\documentclass{article}
\usepackage{amsmath}

\input{tcilatex}

\begin{document}

\title{Nonlinear Equations and wavelets. \\
Multi-Scale Analysis }
\author{A. Ludu, \\
Northwestern State University, Dept. of Chemistry and Physics, \\
Natchitoches, LA 71497, USA \\
R. F. O'Connell and J. P. Draayer\\
Louisiana State University, Dept. Physics and Astronomy, \\
Baton Rouge, LA 70803-4001}
\maketitle

\begin{abstract}
We use a multi-scale similarity analysis which gives specific relations
between the velocity, amplitude and width of localized solutions of
nonlinear differential equations, whose exact solutions are generally
difficult to obtain. 
\end{abstract}

In this paper, wavelet-inspired approaches for localized solutions of NPDE 
are explored [1,2]. A first method provides relations between 
the characteristics of such solutions (amplitude, width and velocity) without
the need of solving the corresponding NPDE. The method uses the
multi-resolution analysis [2] instead of the traditional tools like the
Fourier integrals or linear harmonic analysis which are inadequate for
describing such systems. This scale approach has the advantage that it does
not need the explicit form of the exact solutions. Hence, it is useful
especially in situations when such solutions are unknown. 
The self-similar character of the
fission process of fluid drops is an example where the same type of
singularity occurs in any scale [3,4]. In the following we introduce a
one-scale analysis (OSA) for the NPDE, in terms of their localized traveling
solutions. The so called OSA analysis, described and applied here for
localized traveling solutions belonging to any type of NPDE, provides
algebraic connections between the width $L$, amplitude $A$, and the velocity 
$V$ of the solution, without actually solving the equation. The procedure
consists in the substitution of all the terms in the NPDE, according to the
rules:

i \ \ \ \ \ (substitution valid for traveling waves) $\ \ \ \ \ 
u_{t}\rightarrow -Vu_{x}%
$

ii $\ \ \ \ \ \ \ \ \ \ \ \ \ \ \ \ \ \ \ \ \ \ \ \ \ \ \ \ \ \ \ \ \ \ \ \
\ \ \ \ \ \ \ \ \ \ \ \ \ \ \ \ 
u\rightarrow +A%
$, \ \ \ \ \ \ \ $%
u_{x}%
\rightarrow \pm A/L^{2}$,

and so forth for higher order of derivatives. This substitution in eq.(ii)
is possible only for localized (finite extended support) solutions having at
least one local maximum (like solitons or Gauss functions), if they exist.
Since we are interested in traveling
solutions, the first substitution reduces the number of variables from 2 to
1 so that we are now dealing with an ordinary differential equation instead
of a PDE. Then, the second substitution transforms the ordinary differential
equation into an equation in the parameters describing the amplitude, width
and velocity. Consequently, the NPDE is mapped into an algebraic equation in 
$A,L$ and $V$. The proof of the method follows from the expansion of the
soliton-like solution $u(x,t)=u(s)$ with$s=x-Vt$, in a Gaussian family of
wavelets $\ \Psi (s)=Ne^{Q(s)}$, where $Q(s)$ is a polynomial and $N$ the
normalization constant [2]. If we choose $Q= -i s-\{\{s\symbol{94}2\}%
\TEXTsymbol{\backslash}over 2\}$ we obtain a very particular wavelet with the
support mainly confined in the $(-1,1)$ interval, namely $\Psi (s)=$exp$%
\left[ -is-\frac{s^{2}}{2}\right] $/$\pi ^{1/4}$. We have the discrete
wavelet expansion of $u$\ 

\begin{equation}
u(s)=\sum_{k}C_{j,k}\Psi \left( 2^{j}s-k\right) =\sum_{k}C_{j,k}\Psi
_{j,k}\left( s\right) 
\end{equation}%
in terms of integer translations $k$ of $\Psi $, which provide the analysis
of localization, and in terms of dyadic dilations $2^{j}$ of $\Psi $ which
provide the description of different scales. The idea of the proof is to
choose a point where the expansion in eq.(1) can be well approximated by one
single scale such that $u(s)$ can be approximated with a sum of 
phases in $s$
the OSA approach is given by the convective-dispersive equations, for which
the most celebrated example is provided by the KdV equation

\begin{equation}
u_{t}+uu_{x}+u_{xxx}=0,
\end{equation}%
In Table 1 we
present examples of pure dispersive NPDE, identified in the first column by
the form of the equation, and in the second column by a corresponding
traveling localized solution, if the analytical form is available. Such 
exact solutions provide special relations between $L,A$ and $V$, which are
given in the third column of Table 1. In the last column we introduce the
results of OSA, namely the relations between these three parameters,
provided by eqs.(ii). The usefulness of the approach may be checked, by a
quick comparison between the second and the third columns. The case of
the KdV and MKdV equations, eq(2,3), are described in the first and second
rows of the Table 1, and also are analyzed in previous papers of the same authors. 
Moreover, the same relations like in the KdV case remain valid
even for the solutions of the ''compacton'' type [6]
\begin{equation}
u(x,t)=\frac{\sqrt{32}k\text{cos}\left[ k\left( x-4k^{2}\right) t\right] }{%
3\left( 1-\frac{2}{3}\text{cos}\left[ k\left( x-4k^{2}t\right) ^{2}\right]
\right) },
\end{equation}
where $L=\pi /6k$, that is $L\symbol{126}1/A$, like in the Table 1. Next
example (third row) is provided by a generalised KdV equation, in which the
dispersion term is quadratic
\begin{equation}
u^{t}+(u^{2})_{x}+(u^{2})_{xxx}=0.
\end{equation}%
Eq.(3), known as K(2,2) equation because of the two quadratic terms, admits
compact supported traveling solutions, named compactons [1,5,7-9]. In
general, the compactons are obtained in the form of a power of some
trigonometric function defined only on its half-period, and zero otherwise,
in such a way that the solution is enough smooth for the NPDE in discussion.
In the above example the square of the solution has to be continuous up to
its third derivative with respect to $x$. Different from solitons, the
compacton width is independent of the amplitude and this fact provides the
special connection with the wavelet bases. The compactons are characterize
by a unique scale, and it is this feature that makes it possible to
introduce a nonlinear basis starting from a unique generic function. For
eq.(3) the compacton solution is given by
\begin{equation}
\eta _{c}(x-Vt)=\frac{4V}{3}\text{cos}^{2}\left[ \frac{x-Vt}{4}\right] ,
\end{equation}%
if $|x-Vt|<2\pi $ and $0$ otherwise. Here we notice that the velocity is
proportional to the amplitude and the width of the wave is independent of
the amplitude, $L=4$. As a field of application we mention that the
quadratic dispersion term is characteristic for the dynamics of a chain with
nonlinear coupling. The general compacton solution for eq.(3) is actually a
''dilated'' version of eq.(4). Actually, this combination is just a kink 
compacton joined smoothly
with an antikink one
\begin{eqnarray}
\eta _{KAK}(x-VT;\lambda ) &=& \\
&&0...,  \notag \\
\frac{4V}{3}\cos ^{2}\left[ \frac{x-Vt}{4}\right] \text{ for }-2\pi 
&<&x-Vt<0,  \notag \\
\frac{4V}{3}\text{ for }0 &<&x-Vt<\lambda ,  \notag \\
\frac{4V}{3}\cos ^{2}\left[ \frac{x-Vt-\lambda }{4}\right] \text{ for }%
\lambda  &<&x-Vt<\lambda +2\pi ,  \notag \\
&&0...  \notag
\end{eqnarray}%
Finally, we can construct solutions by placing a compacton on the top
of a KAK. Such a solution exists only for a short interval of time $(\lambda
/V)$, since the two structures have different velocities. The analytic
expression of the solution is given by
\begin{equation}
\eta (x,t)=\eta _{KAK}(x-Vt;\lambda )+\left( \eta _{c}\left( x-V^{\prime
}t-2\pi \right) +\frac{4V}{3}\right) \frac{\chi (x-V^{\prime }t-2\pi )}{2\pi 
},
\end{equation}%
for $0<t<(\lambda -4\pi )/(V^{\prime }-V)$ and zero in the rest. Here $\chi
(x)$ is the support function, equal with 1 for $|x<1$ and $0$ in the rest,
and $V^{\prime }=3$max$\{\eta _{c}/4\}+2V$. For the K(2,2) compacton eq.(4)
fulfills some relations between the parameters: $A=4V/3$ and $L=4$ [7]. The
relations provided by OSA in the last column of the third row, predict such
relations, and hence also proove the existence of the compacton. 
Another good example of the predictive power of the
method is exemplified in the case of a general convection-nonlinear
dispersion equations, denoted by K(n,m)

\begin{equation}
\eta _{t}+\eta _{x}^{n}+\eta _{xxx}^{m}=0,
\end{equation}%
Compacton solution for any $n\TEXTsymbol{\backslash}neq m$ are not known in
general, except for some particular cases. In this case we find a general
relation among the parameters, for any n,m , shown in the fourth and fifth
rows. These general relations L(A,V) approach the known relations for the
exact solutions, in the particular cases like n=m (fourth row), n=m=2 (third
row), n=m=3 (first reference in [1]) and n=3, m=2 ; n=2, m=3 (fifth row).
These results can be used to predict the behavior of solutions for all
values of n,m . 

The situation is different in the case of
compactons, which allow also stationary solutions. When linear and nonlinear
disspersion occur simultaneously, like in the so called K(2,1,2) equation 
\begin{equation}
u_{t}+u_{x}^{2}+u_{xxx}+\epsilon u_{xxx}^{2}=0,
\end{equation}%
where $\epsilon $ is a control parameter, the OSA yields a dependence of the
form $L=\sqrt{\frac{\pm A+\epsilon }{V\pm A}}$, which still provides a
constant width if $V=\pm A+2\epsilon $ . In this case, the speed is
proportional to the amplitude, but can change its sign even at non-zero
amplitude. Solutions with larger amplitude than a critical one $%
(A_{crit}=\mp 2\epsilon )$move to the right, solutions having the critical
amplitude are at rest, and solutions smaller than the critical amplitude
move to the left. This behavior was explored in [7], too. 
A compacton of amplitude A on the top of a infinite-length KAK
solution of amplitude $\delta $
\begin{equation}
u(x,t)=Acos^{2}\left( \frac{x-Vt}{4}\right) +\delta ,
\end{equation}%
is still a solution of the K(2,2) equation, with the velocity given by $V=%
\frac{3}{4}\left( 2\delta +A\right) $. For $A=-2\delta $ the solution
becomes an anti-compacton moving together with the KAK. In the case of a
slow-scale time-dependent amplitude the oscillations in amplitude can
transform into oscillations in the velocity. The key to such a conversion of
oscillations is the coupling between the traditional nonlinear picture
(convection-dispersion-diffusion) and the typical Schrodinger terms. In
Table 2 we present another class of NPDE, namely the dissipative ones. These
equations generalize the linear wave equation (first row) where there is no
typical length of the traveling solutions. The wavelet analysis provides the
correct expression for the dispersion relation $(V=c\rightarrow k^{2}=\omega
^{2}/c^{2})$ with no constraint on either the amplitude $A$ or on the width $%
L$ . In the second row we introduce the Burgers equation which
represents the simplest model for the convective-dissipative interaction.
Dissipative systems are to a large extent indifferent to how they were
initialized, and follow their own intrinsic dynamics. We provide in the
second column an analytic solution of the Burgers equation. For some special
of values of the integration constants $(2C<V^{2},D=0$ $)$the solution
becomes a traveling kink
\begin{equation}
u(x,t)=V+\sqrt{V^{2}-2C}\tanh (\sqrt{V^{2}-2C}(x-Vt)).
\end{equation}%
By applying the OSA approach to the Burger equation (third column) we obtain
the same relation between amplitude and half-width, like in the case of the
exact solution eq.(9), providing the velocity is proportional with the
amplitude. In the following we apply the OSA approach to investigate a
nonlinear Burgers equation
\begin{equation}
u_{t}+au_{x}^{\beta }-\mu u_{xx}^{k}+cu^{\gamma }=0,
\end{equation}%
called quasi-linear parabolic equation [9], and used to describe the flow
of fluids in porous media or the transport of thermal energy in plasma. 
The existence and stability of waves or patterns is strongly dependent on
the coefficients $a,\mu ,c,\beta ,$ and $\gamma $ , and at this point the
OSA can be useful again since there is no general analytic solution for
eq.(10). The result of the OSA approach is presented in the third row of
Table 2. The typical scale of patterns depends on the parameters in the
equations and the amplitude of the excitations, in a complicated way.
However, in order to test OSA again, we found a simple class of exact
solutions when $c=0$ , presented in the fourt row in Table 2, and expressed
as the inverse of a degenerated hypergeometric function. In this expression
we have \ $\mathcal{A}$= $\mu kV^{\alpha -1}F[((k-1)a^{\alpha
})^{-1},z]=(a/V)u^{m-1}$ and $\alpha =(k-1)/(m-1)$. The asymptotic behavior
of the left hand side of the solution given in fourth row, second column, is
described by 
\begin{equation}
\Gamma (\alpha -1)\left[ \left( -1\right) ^{\alpha }+\frac{1}{\Gamma (\alpha
)}z^{\left( \alpha -1\right) }e^{\alpha }\right] +\mathcal{O}(1/z).
\end{equation}%
If z approaches + $\infty $ the solution increases indefinitely like an
exponential. For $\alpha $ \TEXTsymbol{>}1 (strong difussion effects), for
even k and for even $\beta $ , the traveling wave $u(x-Vt)$ has a negative
singularity towards -$\infty $ at $x+x_{0}=\Gamma (\alpha +1)(-1)^{\alpha }<0
$ . For $k$ odd there is also a singularity at $x+x_{0}>0$ . 
If $k$ is even and $\beta 
$ is odd (the singularity is pushed towards imaginary $x$ ), or if $0<\alpha
<1$ , the singularity is eliminated and the solution becomes semi-bounded,
like in the particular situations investigated in the article [9]. In this
case, OSA provides again the correct relations, since we obtaine the special
behavior of the solution if the velocity is proportional to the power $\beta
-1$ of the amplitude $A$ . Also, we predict the space scale of these
semi-compact pulses, namely the length $L=\frac{\mu k^{2}A^{k-1}}{V\pm
a\beta A^{m-1}}$ . The OSA analysis can be applied in the case of
sine-Gordon equation, fifth row of Table 2. The solutions with the velocity
proportional with $L^{2}$ are characterized through the OSA approach by a
transcendental equation in A, identical with the equation fulfilled by the
amplitude A of the exact sine-Gordon soliton. In the sixth row, we present
the cubic nonlinear Schrodinger equation (NLS3) which has a soliton
solution. 
In the sixth row of Table 2 we present the NLS3 equation
together with its one soliton solution of amplitude $\eta _{0}$ , obtained
by the inverse scattering method. In the last column we also show the
relation between the parameters of a localized solution, obtained by OSA.
The equation for $L(A,V)$ is more general than that one fulfilled by the
soliton, and hence is related to more general localized solutions. By
choosing the velocity proportional to the amplitude, we reobtain the $L\sim
1/A=1/\eta _{0}\sim 1/V$ typical relations for the soliton given in the
second column. 
\begin{equation}
-\frac{\hbar ^{2}}{2m}\Psi _{xx}+(E-V)\Psi +a\Psi ^{3}=0,
\end{equation}%
then the $L$ parameter gives an estimation for the wavelength of the
wavefunctions, or for the correlation length in a Bose model
\begin{equation}
L=\frac{\hbar }{\sqrt{2m(E-V)+aA^{2}}}\simeq \frac{\hbar }{\sqrt{2m(E-V)}}%
\text{for small }A.
\end{equation}%
For the general case of a NLS equation of order $n$ (seventh row), where a
general analytical solution is unknown, the method predicts a special $%
L=L(A,V)$ dependence, shown in the third column. Contrary to third order
NLS, where the dependence of L with A is monotonous for $V=\sim \pm A$ $(n=3)
$, at higher orders than 3, the $L(A)$ function has discontinuities in the
first derivative. This wiggle of the function $n=4$ hold at a critical
width, possibly producing bifurcations in the solutions and scales. As a
consequence, initial data close to this width can split into doublet (or
even triplet, for higher order NLS) solutions, with different amplitudes.
Such phenomena have been put into evidence in several numerical experiments
for quintic nonlinear equations [8-10]. The final example of Table 2 is
provided by the Gross-Pitaevski (GP) mean field equation, which is used to
describe the dilute Bose condensate [11]. The scalar field (or order
parameter) governed by this equation was shown to behave in a particle
manner, too, since it can contain topological deffects, namely dark
solitons. The space scale L of such solutions is important, for both the
theory and experiment, since is related to the trap dimensions and to the
scattering length. In the last row of the Table 2 we give one particular
solution of a simplified one-dimensional version of the GP equation [12]
\begin{equation}
i\hbar \frac{\partial \Psi (\overrightarrow{x},t)}{\partial t}=\left( \frac{%
-\hbar ^{2}}{2m}\triangle +V_{ext}(\overrightarrow{x})+\frac{4\pi \hbar ^{2}a%
}{m}\Psi (\overrightarrow{x},t)^{2}\right) \Psi (\overrightarrow{x},t),
\end{equation}%
where $a$ is the s-wave scattering length and $V_{ext}$ is the confining
potential. In the solution provided in the table, the half-width of the
exact nonstationary solution is $L=1/\sqrt{\nu _{c}^{2}-p^{2}}$, where $%
v_{c}=\sqrt{1-aV}$ is the Landau critical velocity, and $p=dq(t)/dt$ is the
momentum associated with the motion of this disturbance. It is easy to check
that the OSA provides a good match with this exact solution, and also L fits
the correlation length $l_{0}=\sqrt{\frac{m}{4\pi \hbar ^{2}a}}$ . We stress
that such estimation of the length is also important in nuclear physics
where one can explain the fragmentation process as a bosonization in 
$\alpha$-particles, inside the nucleus. Such systems
are coherent if the wavelength associated with the cluster (the resulting L
in the GP equation) is comparable with the distance between the $\alpha $
-clusters. 
\begin{equation}
u_{t}+au_{x}^{m}+\mu u_{xx}^{k}+cu_{xxx}^{n}=0.
\end{equation}%
Here $m,k$ and $n$ are integers and the corresponding terms are responsible
of the nonlinear interaction (convective term), dissipation and dispersion
[9]. The above equation is related to weakly nonlinear phenomena, and it
occurs in modeling porous medium, magma, interfacial phenomena in fluids
(and hence applications to drop physics), etc. Thi OSA approach maps this equation into
\begin{equation}
amA^{m-1}-VL^{2}-\mu k^{2}A^{k-1}L+n^{3}A^{n-1}=0,
\end{equation}%
Table 3, first row. The most symmetric case is obtained when either $V=0$
(stationary patterns) or $V\sim A^{m-1}$ . In this situation the condition
to have a monotonous dependence of \ $L$ as a function of $A$ is $2k=m+n$
which yields a scale structure
\begin{equation}
L=A^{k-m}\left( \frac{\mu k^{2}\pm \sqrt{\mu ^{2}k^{4}-4mn^{3}(a-V_{0})}}{%
2m(a-V_{0})}\right) \sim A^{k-m},
\end{equation}%
where we put $V=mV_{0}A^{m-1}$ . The condition $2k=m+n$ is just the
condition obtained in [9] from a scaling approach. This condition assures
the universality of the corresponding patterns, and it is the unique case in
which $L$ depends on a power of $A$. In the above cited paper, the author finds out the
condition for mass invariance at scaling transformations as $m=n+2=k+1$. In
our case we just have to request the product $AL$ (which gives a measure of
the mass, or volume of the pattern, like in the case of one-dimensional
solitons) to be a constant. This gives the condition $k-m+1=0$ which,
together with the general invariance condition $2k=m+n$ , reproduces $%
m=n+2=k+1$ . In this case we have patterns characterized by a width
\begin{equation}
L=\frac{\mu k^{2}\pm \sqrt{\mu ^{2}k^{4}-4mn^{3}(a-V_{0})}}{2mA(a-V_{0})}%
\rightarrow \frac{n^{3}}{A\mu k^{2}}.
\end{equation}%
If $a\sim V_{0}$ the width approaches $n^{3}/A\mu k^{2}$ . In order to make $%
L$ independen of $A$ , like in the compacton case, we need $m=k$, which
together with the first invariance condition $2k=m+n$, yields $m=n=k$. This
is the exceptional case when the dissipative and dispersive processes have
the same scaling, resulting form the invariance of the eq.(12) under the
group of scales. Finaly, if we choose $L\sim V$ we obtain the condition $%
k+1=2m$ which (together with $2k=m+n$) is the condition for spiral symmetry
and occurence of similarity structures [9]. The next example is provided by one 
of the most
generalized KdV equation, which is generated from the Lagrangian [5]
\begin{equation}
\mathcal{L}(n,l,m,p)=\int \left[ \frac{\phi _{x}\phi _{t}}{2}+\alpha \frac{%
\phi _{x}^{p+2}}{(p+1)(p+2)}-\beta \phi _{x}^{m}\phi _{xx}^{2}+\frac{\gamma 
}{2}\phi _{x}^{n}\phi _{xx}^{l}\phi _{xxx}^{2}\right] dx,
\end{equation}%
where $\alpha $, $\beta $ and $\gamma $ are parameters adjusting the
relative strentgh of the interactions, and $n$, $l$, $m$, $p$ are integers.
For example, for $\gamma =0$ one re-obtains the K(2,2) equation, and for $%
\gamma =m=0,p=1$ one obtains the KdV equation. The associated Euler-Lagrange
equation in the function $\phi _{x}=u(x,t)\rightarrow $ $u(x-Vt)=u(y)$,
reads after one integration
\begin{eqnarray}
Vu &=&\frac{\alpha }{p+1}u^{p+1}-\beta mu^{m-1}u_{y}^{2}+2\beta
(u^{m}u_{y})_{y}+\frac{\gamma n}{2}u^{n-1}(u_{y})^{l}(u_{yy})^{2} \\
&&-\frac{\gamma l}{2}(u^{n}(u_{y})^{l-1}(u_{yy})^{2})_{y}+\gamma
(u^{n}(u_{y})^{l}u_{yy})_{yy}+C,  \notag
\end{eqnarray}%
where $C$ is the integration constant. By using the OSA we obtain the
following important result, expressed in the second row of Table 3: The
unique case when such an equation allows compact supported traveling
solutions is when $m=p=n+r$ , $C=0$ and $V=V_{0}A^{m}$ . This result is in
full agrement with the variational calculation in [5]. Both eqs.(12) and
(16) are rather more qualitative than capable of modeling measurable
phenomena. That is why we introduce now a more general model equation, in
the form
\begin{equation}
u_{t}+f(u)_{x}+g(u)_{xx}+h(u)_{xxx}=0,
\end{equation}%
where $f,g$ and $h$ are differentiable functions of the the function $u(x,t)$
itself. The OSA approach gives the equation
\begin{equation}
-V+f^{\prime }(A)+\frac{Ag^{\prime \prime }(A)+g^{\prime }(A)}{L}+\frac{%
A^{2}h^{\prime \prime \prime }(A)+3Ah^{\prime \prime }(A)+h^{\prime }(A)}{%
L^{2}}.
\end{equation}%
A general analysis of eq.(17) is difficult, and the best ways are numerical
investigations obtained for particular choices of the three functions. We
confine ourselves here only to show that the class of solutions which have
similarity properties are those for which $V=V_{0}f^{\prime }(A)$ . In this
case eq.(18) can be reduced to
\begin{equation}
L^{2}f^{\prime }(1-V_{0})+L(Ag^{\prime \prime }+g^{\prime })+A^{2}h^{\prime
\prime \prime }+3Ah^{\prime \prime }+h^{\prime }=0,
\end{equation}%
case which is presented in the third row of Table 3. This last relation can
be used for different purposes. For example, given a certain type of
dispersion and difusion ( $g,h$ fixed), we can estimate for what types of
nonlinearity ( $f$) the width L will have a given dependence with $A$ . Or,
if we know for instance $f(u)=f_{0}u^{q_{1}}$ and $h(u)=h_{0}u^{q_{2}}$ , we
can ask what type of diffusion $g$ \ we need, to have constant scale (width)
of the patterns (waves), no matter of the magnitude of the amplitude $A$ .
In other words, which is the compatible diffusion term, for given
nonlinearity-dispersion terms, which provides fixed scale solutions. The
result is obtained by integration eq.(19) with respect to $g(u)$
\begin{equation}
g(u)=-\frac{h_{0}}{L}\left( 1+q^{2}+\frac{1}{q^{2}-1}\right) u^{q_{2}}-\frac{%
Lf_{0}(1-V_{0})}{q_{1}-1}u^{q_{1}}+C_{3}\text{Log }u+C_{4,}
\end{equation}%
where $C_{3,4}$ are constants of integration. In a similar way one can check
the existence of different other configurations by solving eq.(19), or more
general, eq.(18). A last application of this method, occurs if the KdV
equation has an additional term depending on the square of the curvature
\begin{equation}
u_{t}+u_{x}+u_{xxx}+\epsilon \left( u_{xx}^{2}\right) _{x}=0.
\end{equation}%
This is the case for extremely sharp surfaces (surface waves in solids or
granular materials) when the hydrodynamic surface pressure cannot be
linearized in curvature. Such a new term yields a new type of localized
solution fulfilling the relations
\begin{equation}
L=\sqrt{\frac{4\epsilon A}{-1\pm \sqrt{1-8\epsilon A(A\pm V)}}}.
\end{equation}%
If we look for a constant half-width solution (compacton of $1/L=\alpha $)
we need a dependence of velocity of the form $V=(1+\alpha ^{2}\epsilon
/8)A+1/8\epsilon A+\alpha /4$ . There are many new effects in this
situation. The non-monoton dependence of the speed on $A$ introduces again
bifurcations of a unique pulse in dublets and triplets. Also, there is an
upper bound for the amplitude at some critical values of the width. Pulses
narrower than this critical width drop to zero. Such bumps can exist in
pairs of identical amplitude at different widths. They may be related with
the recent observed ''oscillations'' in granular materials [7]. The
examples presented in Tables 1-3 prove that the above method provides a
reliable criterion for finding compact suported solutions. 

\bigskip

REFERENCES

1. P. Rosenau and J. M. Hyman, \textit{Phys. Rev. Let}. \textbf{70} (1993)
564; B. Dey,\textit{\ Phys. Rev. E} \textbf{57} (1998) 4733.

2. I. Doubechies and A. Grossmann, \textit{J. Math. Phys.} \textbf{21}
(1980) 2080; A. Grossman and J. Morlet, \textit{SIAM J. Math. Anal.} \textbf{%
15} (1984) 72;

3. \ A. Ludu and J. P. Draayer, \textit{Phys. Rev. Lett.} \textbf{10}
(1998) 2125; J. M. Lina and M. Mayrand, \textit{Phys. Rev. E} \textbf{48}
(1994) R4160.

4. J. L. Bona, \textit{et al}, \textit{Contemp. Math.} \textbf{200} (1996)
17 and \textit{Phil. Trans. R. Soc. Lond. A} \textbf{351} (1995) 107; G.
Zimmermann, \textit{Proceedings Int. Conf. on Group Theoretical Methods in
Physics, \ G22} (Hobart, 10-18 July, 1998) and private communication.

5. F. Cooper, J. M. Hyman and A. Khare, \textit{Compacton Solutions in a
Class of Generalized Fifth Order KdV Equations} in press.

6. C. N. Kumar and P. K. Panigrahi, Preprint \textbf{solv-int/9904020}.

7. P. Rosenau, \textit{Phys. Lett. A} \textbf{211} (1996) 265 and \textit{%
Phys. Rev. Lett} \textbf{73} (1994) 173.

8. J. M. Hyman and P. Rosenau, \textit{Physica D} \textbf{123} (1999) 502.

9. P. Rosenau, \textit{Physica D} \textbf{123} (1999) 525; \textbf{8D}
(1983) 273.

10. V. G. Kartavenko, A. Ludu, A. Sandulescu and W. Greiner, \textit{Int. J.
Mod. Phys. E} \textbf{5} (1996) 329.

11. M. H. Anderson \textit{et. al., Science} \textbf{269} (1995) 198.

12. Th. Busch and J. R. Anglin, \textit{Phys. Rev. Lett.}, \textbf{84}
(2000) 2298.

\begin{table}[t]
\caption{Traveling localized solutions for nonlinear dispersive equations.\label{tab:exp}}
\vspace{0.2cm}
\begin{center}
\begin{tabular}{cccc}
\hline \\
NPDE & Analytic solution and the && OSA \\
&relations among parameters & &approach\\
\hline \\

$u_{t}+6uu_{x}+u_{xxx}=0$ & $A~\hbox{sech} ^2 {{x-Vt} \over L};$ \ \ \  $L=\sqrt{2 / A}$, && $L=|V\pm 6A|^{-1/2}$\\
 &  $V=2A$ &&If $V\sim A$, $L\sim A^{-1/2}$ \\ 
\hline \\
$u_{t}+u^2u_{x}+u_{xxx}=0$ &   $A~\hbox{sech } {{x-Vt} \over L};$ \ \ \  $L=1/A$, & &
$L=|V \pm 6A^2 |^{-1/2}$ 
\\
&  $A=\sqrt{V}$  &&If $V\sim A^2 ,L\sim A^{-1}$\\
\hline \\
$u_{t}+(u^2)_{x}+(u^2
)_{xxx}=0$  &  $A \hbox{cos} ^2 {{x-Vt}\over L}$, \ \ \hbox{if} \ \ 
$|(x-Vt)/4|\leq \pi /2 $;
 & & $L=\biggl ( {{8A}\over
{|V
\pm 2A|}}\biggr ) ^{1/2}$  \\  
 & L=4 &&   \\
\hline \\
$u_t +(u^n )_x +(u^n )_{xxx}=0$  & $\biggl [ A cos^{2}\biggl ( {{x-Vt} \over {L}}
\biggr ) \biggr ] ^{1 \over {n-1}}$, \ \  if $|x-Vt| \leq {{2n\pi}\over{n-1}}$ & &
$L=\biggl ( {{n(n^2 +1)}\over{\alpha \pm n}}\biggr )^{1/2}$ \\
&and 0 else; & & \\
&$L={{4n}\over {(n-1)}},$ \ \ $A={{2Vn}\over {n+1}} $ && if $V=\alpha A^{n-1}$ \\ 
\hline \\
 $u_t +(u^n )_x +(u^m )_{xxx}=0$ & unknown 
&  & 
$L=\biggl ({{{n(n^2 +1) A^{n-1}} \over {V \pm mA^{m-1}}}}\biggr )^{1/2}$
\\  
$n\neq m$ &in general &
 & \\ 
\hline \\
\end{tabular}
\end{center}
\end{table}

\begin{table}[t]
\caption{Traveling localized solutions for nonlinear diffusive equations.\label{tab:exp}}
\vspace{0.2cm}
\begin{center}
\begin{tabular}{cccc}
\hline \\
NPDE & Analytic solution and the & OSA \\
&relations among parameters & approach\\
\hline \\
$u_{xx}-{1 \over  {c^2}} u_{tt}=0$ & $\sum C_k e^{i(kx\pm \omega t )}$;  & 
$V=c$ \\
 &$k^2 = \omega ^2 /c^2$ & $ A, \ L$   arbitrary  \\
\hline \\
$u_t + uu_x - u_{xx}=0$ & $ \sqrt{C -V^2} \hbox{tan}(\sqrt{C -V^2}{{x-Vt}\over {2}} +D)$
 & $L=(A \pm V)^{-1}$ \\
 & $+V$ & If $V\sim A$,  $L\sim 1/A$\\
\hline \\
$u_t +a (u^m )_x -\mu  (u^k )_{xx}+cu^{\gamma}$ & only particular cases
\ \    & $cA^{\gamma} L^2 + (V\pm amA^{m-1})L$ \\
=0 &known & $\pm \mu k^2 A^{k-1}=0$   \\
\hline \\
$u_t +a (u^m )_x -\mu  (u^k )_{xx}=0$ & $ -{\cal A} z^{\alpha }\ _{1}F_{1}(\alpha , \alpha +1 ,
z) = x+x_0$
\ \    &$L={{\mu k^2}\over{am-\alpha }}A^{k-m}$, \\
 & & if $ V=\alpha A^{m-1}$   \\
\hline \\
$u_{xt}-\sin u=0$ &  $A~\tan^{-1}\gamma ~e^{{x-Vt}\over {L}}$ \ \ \  &
$\pm {{VA}\over {L^2 }}=sin A$ \\ 
& &    If $V=L^2, A=sinA$ \\ 
\hline \\
$i\Psi_{t}+\Psi _{xx}+2|\Psi |^2 \Psi =0$  & $\eta_0 e^{i(\omega t + kx)} sech[ \eta_0 
(x-Vt)]$;
\ \   &  $L={{\pm V \pm \sqrt{|V^2 - 4 A^2 |}}\over {2A^2 }}$ \\
&$ L=1/ \eta_0$&  If $A\sim V, \ \ L=1/A$ & \\ 
\hline \\
$i\Psi_{t}+\Psi _{xx}+|\Psi| ^{n-1}\Psi=0$  & unknown in general    
&$L={{\pm V \pm \sqrt{|V^2 - 4 A^n |}}\over {2A^n }}$ \\
 & &   \\ 
\hline
$i\Psi_t=-{1 \over 2}\triangle \Psi $ &
$ip+\sqrt{v_{c}^{2}-p^2}\times  $  & 
$L=(aA^2 \pm V -1)^{-1/2} $\\
$+[a|\Psi|^2 +V(x) -1]\Psi$  & $\hbox{tanh}[a\sqrt{v_{c}^{2}-p^2}(x-q(t) )]$ 
& If $V\sim \pm 1, L\sim 1/(A\sqrt{a})$ \\ 
\hline
\end{tabular}
\end{center}
\end{table}

\begin{table}[t]
\caption{Traveling localized solutions for  dissipative-dispersive equations.\label{tab:exp}}
\vspace{0.2cm}
\begin{center}
\begin{tabular}{cccc}
\hline \\
The NPDE equation & OSA approach \\
\hline \\

$u_t + a(u^m ) _{x} + b (u^k )_{xx}+ c(u^n)_{xxx}=0$; & 
$L=A^{m-k} \cdot {{\mu k^2 \pm \sqrt{\mu ^2 k^4 -4mn^3
(a-V_0 )}}\over {2m(a-V_0 )}}$ \\
&if  $V=mV_0A^{k-1}$  \\
\hline \\
$Vu={{\alpha }\over {p+1}}u^{p+1}-\beta mu^{m-1}(u_y )^2 +2\beta (u^{m} u_y )_{y}$ &  
$2L^{l+4}((n+l+1)V_0 -\alpha) $\\ 
$+{{\gamma n}\over {2}} u^{n-1} (u_y )^l (u_{yy})^ 2 -{{\gamma l} \over 2} (u^{n} (u_y
)^{l-1}(u_{yy})^2 )_{y}
$&$-2L^{l+2}(l+n+1)(l+n+2)\beta$ \\ 
$+\gamma (u^n (u_y )^l u_{yy})_{yy} +C$& $-(l+n+1)(2+2n^2 +3l+l^2 +n(5+3l))\gamma$ \\ 
&if $C=0$, $V=V_0 A^{m}$ and $m=p=n+l$ \\
\hline \\
$u_t +f(u)_x +g(u)_{xx}+h(u)_{xxx}=0$  &  $L=-\biggl [ g'+Ag''\mp \biggl ( (Ag''+g')^2 $ \\ 
& $ -4f'(1-V_0 )(A^2 h'''+3Ah''+h') \biggr ) ^{1/2} \biggr ] $ \\
& $ \times (2A^2 h'''+6Ah''+2h')^{-1}$ \\ 
& if  $V=V_0 f'(A)$ \\
\hline \\
\end{tabular}
\end{center}
\end{table}

\end{document}